# High-order multipoles in all-dielectric metagrating enabling ultralarge-angle light bending with unity efficiency

*Tie-Jun Huang, Li-Zheng Yin, Jin Zhao, Chao-Hai Du and Pu-Kun Liu*[*]

State Key Laboratory of Advanced Optical Communication Systems and Networks, Department of Electronics, Peking University, Beijing, 100871, China

**ABSTRACT**: Gradient metasurfaces have been extensively applied in recent years for enabling an unprecedented control of light beam over thin optical components. However, these metasurfaces suffer from low efficiency when it comes to bending light with large angle and high fabrication demand when it requires fine discretion. In this work, we investigate the all-dielectric metagrating based on mie-type resonances interference, allowing extraordinary optical diffraction for beam steering with ultralarge angle. It is found that the coupling inside and among lattice of metagrating can tune the exciting state of electric and magnetic resonances including both fundamental dipoles and high-order multipoles, leading to ideal asymmetrical scattering pattern for redistributing the energy between the diffraction channels at will. The participation of quadrupole and hexapole not only significantly enhance the working efficiency, but also bring distinctive possibilities for wave manipulation which cannot be reached by dipoles. Utilizing a thin array of silicon rods, large-angle negative refraction and reflection are demonstrated with almost unity efficiency under both polarizations. Compared with conventional metasurfaces, such an all-dielectric mategrating has the merits of high flexibility, high efficiency and low fabrication demand. The strong coupling and prosperous interactions among multipoles may behave as a cornerstone for broad range of wavefront control and offer an effective solution for various on-chip optical wave control such as bending, focusing, filtering and sensing.

The rapid development of exotic metamaterials has emerged unprecedented manipulation of electromagnetic waves which cannot be attained by natural materials.[1,2] As one of the most promising concepts, the metasurface by imparting abrupt optical properties through subwavelength elements, has attracted tremendous attention in optics community since it was introduced by the group of Capasso in 2011.[3] Theoretically, metasurfaces can freely modulate the basic features of light including its amplitude, phase and polarization, then enabling unprecedented wave control.[4-6] With the property of ultra-thin thickness, the metasurface has been applied in versatile intriguing applications, such as anomalous beam manipulation,[7] focusing lens,[8,9] nodiffraction-limited waveguide,[10] broadband absorber,[11] optical vortex generation,[12] holography[13] and so on. In general, tailoring incident waves strongly relies on accumulating the gradient changes of periodic elements,[4-7] i.e., the variation of geometric size or rotating angle. To achieve enough resolution in the control process, the fine spatial discretization and elaborate fabrication are demanded. For the large-angle beam routing, owing to the the intrinsic drawbacks of gradient metasurfaces,[14,15] complex geometric configurations are needed to avoid the dramatical decrease of efficiency.[16] The corresponding manufacture resolution will have considerable influence on the working performance of optical devices. In addition, the plasmonic nanoantennas with the ability of controlling light in subwavelength scale are widely employed in designing metasurfaces at infrared and optical wavelengths, but the inherent high loss leads to further decrease of working efficiency.[1-4]

Recently, owing to the low-dissipative loss, high-permittivity dielectric nanoparticles which support mie-type resonances have arisen as favorable candidates for various nanophotonics applications.[17-19] Compared with their metallic counterparts, the dielectric nanoparticles are able to excite a rich phenomenology of strong electric and magnetic mutilpoles without using the complex geometries such as split-ring resonators and U-shaped particles.[20] The low-loss performance in combination with the localized multipolar nature can provide resonances with high quality and serval orders of magnitude field enhancement.[21,22] Besides, the resonances are sustained inside the dielectric particles,[23] unlike the plasmonic structures which mainly rely on surface modes. These merits contribute to strong light-matter interaction and are promising for nonlinear phenomena,[23,24] Fano resonances,[25,26] sensing,[27] guiding,[28] spectroscopy,[29] magnetic mirrors,[30] and so on. Most importantly, by tuning the shape or geometric parameters of dielectric particles, the interplay among multiple electric and magnetic multipoles can produce great opportunities for unconventional scattering control.[17,18] For instance, by overlapping the electric dipole (ED) and magnetic dipole (HD) with identical values, the no-backward or no-forward scattering can be realized, which is also known as the Kerker effect.[31] The structures including nanospheres,[31,39] nanocubes,[30,32] nanodisks,[28,33-35] nanorods,[36] oligomers,[24,25] and C-shaped particles[37] have been applied in forming the directional scattering patterns, the functionalities of which are extended into photoluminescence emission,[34] abosorb,[38] light harvest,[39] artificial colors[32,35] and so on. What should be pointed out that the directional scattering property based on interference between electric and magnetic multipoles is the essential condition for building Huygens' metasurfaces with unity working efficiency and a $2\pi$ phase shift.[30,36,37,40-43] However, the Huygens' metasurfaces are also plagued by the high-demand fabrication and limited working performance in large bending angles. The demonstrated anomalous refractive angles of Huygens' metasurfaces are often smaller than 30 degree.[40-43]

Diffraction effect, induced by period structures, can transfer the incident energy into a discrete set of diffraction channels according to the Floquet theory. Under this circumstance, a large bending angle of light is formed by uniform element whose size does not demand severe fabrication resolution. The metasurface adopting the diffraction effect is also known as metagrating which can achieve extreme wave manipulation.[44-46] The very essential principle for the metagrating is completely suppressing all unselected diffraction channels using asymmetrical scattering element. This suppression is still difficult to be realized, owing to each channel is not independently controllable.[47] The interplay of mie resonances with defined multipolar characters provides a novel platform for forming scattering pattern and choosing the specified diffraction order. Recently, utilizing simple dielectric grating configurations, some pioneering researches have achieved promising results in anomalous refraction with angle up to 90 degree.[48-53] However, there remain two important issues plaguing the development of dielectric metagrating based on mie-type resonances. Firstly, the influence of interaction between nanoparticle is comparable to the effect of the resonance inside individual element.[54] Therefore, the scattering behavior of periodic structures can be totally different from isolated lattice.[55] Some studies also utilize the lattice coupling to overlap the spectral of ED and MD, but the related designs for abnormal wavefront control have not yet been investigated.[54] Many other investigations just neglect the influence of lattice coupling[48,56] or discuss the coupling insufficiently.[49-53] Secondly, except the fundamental ED and MD resonances, the high-oder multipolar modes can offer new possibilities for scattering pattern control and complex wave manipulation.[57] Although some works have realized the importance of these modes by enhancing scattering directionality or building magnetic mirrors,[55,58,59] the method of applying and optimizing high-order multipoles in diffraction selection is still lacked.

In this work, we analytically investigate the efficient wave manipulating with large angle induced by resonant multipoles insides periodic high-index nanoparticles. Based on the multiple scattering theory, both the simple grating (SG) formed by a single dielectric rod in one period and the compound grating (CG) formed by a dimer in one period are investigated in detail. It is found the coupling effect induced by the grating configurations can take not only electric and magnetic dipoles but also high-order multipoles into waves manipulation. With appropriate designs, the scattering properties of the nanoparticle are elaborately tailored through the interference between multipoles, allowing harnessing light energy in both the transmission and reflection diffraction channels. In this process, the high-order multipoles can not only help to improve the working performances formed by dipoles, but also play dominant roles for wave control and exhibit different features with dipoles, leading to the enhancement of flexibility and working efficiency. On the basis of this, efficiencies close to 100% for large-angle negative reflection and reflection of incident waves are specifically demonstrated for both TM and TE polarizations. The low-loss all-dielectric structure may render a new strong impetus for designing compact optical components with low profile and high efficiency.

**RESULTS AND DISCUSSION**

Let us start by considering one-dimensional rods array with refraction index being 3.5 (approaching silicon). As shown in **Figure** 1a and b, the lattice of the array is composed of one or two dielectric cylinders, which is denoted as the SG or CG. The period of the lattice is $L$ and cylinders are uniform in the z direction. For the SG case, all the rods have the same radius $r$. For the CG case, two cylinders with radiuses being $r_1$ and $r_2$, has a center-to-center distance $d$. In the local coordinate system of one specified lattice, the line connecting two rods has an angle $\phi$ with respect to the x-axis. To simplify the deduction, we firstly consider two-dimensional scattering cases when a plane wave with incident angle $\theta$ impinges on the grating. In the last section, the realizable three-dimensional cases are also calculated and discussed. The polarization of incident waves can be TM or TE when the electric or magnetic field is polarized along z-axis. Under polar coordinate $(\rho, \varphi)$, the scattering field of an isolated cylinder is expanded into a series of cylindrical harmonics, as expressed by

$$F = \mathbf{P}^T \cdot \mathbf{\sigma} \tag{1}$$

where $\mathbf{P} = [i^m H_m^{(1)}(k_0\rho)e^{im\varphi}]$ and $\mathbf{\sigma} = [\sigma_m]$ are column vectors. $F$ can be $E_z$ or $H_z$ component for the TE or TM polarization. $H_m^{(1)}$ is $m$-th order Hankel function of the first kind and $\sigma_m$ is the normalized[66] (by $2\lambda/\pi$, $\lambda$ being the working wavelength) scattering coefficient of the $m$-th cylinder harmonic with the form of

$$\sigma_m^{TE} = \frac{J_m(Nk_0r)J'_m(k_0r) - NJ_m(k_0r)J'_m(Nk_0r)}{NJ'_m(Nk_0r)H_m^{(1)}(k_0r) - J_m(Nk_0r)H'^{(1)}_m(k_0r)} \tag{2a}$$

$$\sigma_m^{TM} = \frac{NJ_m(Nk_0r)J'_m(k_0r) - J_m(k_0r)J'_m(Nk_0r)}{J'_m(Nk_0r)H_m^{(1)}(k_0r) - NJ_m(Nk_0r)H'^{(1)}_m(k_0r)} \quad (2b)$$

Here, $J_m$ is the $m$-th order Bessel function, $J'_m$ and $H'^{(1)}_m$ is the derivatives of the original functions, $N$ is the refractive index of the rod. Then, for a single rod, its total scattering cross section $\sigma_s$ is the synthesis of all scattering orders

$$\sigma_s = \sum_m |\sigma_m|^2 \quad (3)$$

For TM polarized waves, the cases of $|m| = 0, 1, 2, 3$ are corresponding to the MD, ED, electric quadrupole (EQ) and electric hexapole (EH). The cases of $|m| = 0, 1, 2, 3$ for TE polarization are related with ED, MD, magnetic quadrupole (MQ), and magnetic hexapole (MH). The scattering spectrums of an isolated silicon rod under two polarizations are plotted in Figure S1a and b (section 1 of the supplement). The field distributions and scattering properties of these multipoles also calculated and illustrated in the supplement. It is clear that, the electric and magnetic multipoles exist under both polarizations, significantly enriching the scattering phenomenon of the dielectric rod. Note that, owing to the feature of rotational symmetry, the opposite orders always hold the same scattering coefficient and a certain phase difference, i.e., $\sigma_m = \sigma_{-m}$.

When the cylinders form a one-dimensional SG, the detailed scattering properties can be analytically investigated by the multiple scattering theory. Then, the scattering field of the SG is expressed by[60]

$$F^{SG} = \sum_{l=-\infty}^{\infty} \mathbf{P}_l^T \cdot \boldsymbol{\sigma}^{SG} e^{-ilk_0 L\cos\theta} \quad (4)$$

Here, $\boldsymbol{\sigma}^{SG} = [\sigma_m^{SG}]$ and $\mathbf{P}_l = [H_m^{(1)}(k_0\rho_l)e^{im\varphi_l}]$ are column vectors. $\sigma_m^{SG}$ is the unknown scattering coefficient of $m$-th order multipole and $\hat{\boldsymbol{\rho}}_l = [\boldsymbol{\rho} - lL\hat{\mathbf{x}}] = \rho_l(\hat{\mathbf{x}}\cos\varphi_l + \hat{\mathbf{y}}\cos\varphi_l)$. Using the iterative T-matrix,[60] the scattering coefficients can be retrieved by the following relation[61]

$$\left(\bar{\mathbf{I}} - \bar{\mathbf{T}} \cdot \bar{\mathbf{C}}\right) \boldsymbol{\sigma}^{SG} = \bar{\mathbf{T}} \cdot \mathbf{B} \tag{5}$$

Here $\mathbf{B} = [(-i)^n e^{in\theta}]$, $\bar{\mathbf{I}}$ is the unit matrix, $\bar{\mathbf{T}}$ is a diagonal matrix with $T_{mm} = \sigma_m$, $\bar{\mathbf{C}} = [C_{ml}]$ is the matrix of lattice sum with

$$C_{ml} = \left[\sum_{g=1}^{\infty} H_{m-l}(k_o gL)\left(e^{-imk_0 L\cos\theta} + (-1)^{m-l} e^{imk_0 L\cos\theta}\right)\right] \tag{6}$$

which can be calculated by an integral process. Clearly, the scattering phenomenon of a cylinders within a SG is also expressed by cylindrical harmonics with coefficient $\sigma_m^{SG}$. Note that, the lattice coupling can lead to multiple intro-coupling among all the multipoles, which makes the cylindrical harmonics with opposite sign no longer behave as a whole, i.e., $\sigma_m^{SG} \neq \sigma_{-m}^{SG}$. For the CG case, two rods in each unit cell of CG are considered as a single composite structure, the scattering features of which are also described by the multipoles with coefficient $\boldsymbol{\sigma}^{CG} = [\sigma_m^{CG}]$. Then, the theoretical deduction of CG is similar to the above, except that the T-matrix should consider the coupling effect inside the unit cell. The scattering waves from one rod would become the incident waves for another rod. After considering this, the new **T** matrix $\bar{\mathbf{T}} = [\bar{t}_{mn}]$ for CG is rewritten as

$$\bar{\mathbf{T}} = \bar{\mathbf{T}}_1 \cdot \bar{\boldsymbol{\beta}}_{01} + \bar{\mathbf{T}}_2 \cdot \bar{\boldsymbol{\beta}}_{02} \tag{7}$$

where $\bar{\mathbf{T}}_1$ and $\bar{\mathbf{T}}_1$ are the **T** matrix of each rod under their coordinates, $\bar{\boldsymbol{\beta}}_{01}$ and $\bar{\boldsymbol{\beta}}_{02}$ are the transform matrixes between different coordinates. The elements of $\bar{\mathbf{T}}$ describe the couplings between two rods in a lattice, including electric-electric couplings, magnetic-magnetic couplings and electric-magnetic couplings. The deduction of equation 7 and other details of $\bar{\mathbf{T}}$ are put in section 2 of the supplement. By submitting the composite $\bar{\mathbf{T}}$ into equation (5), the effective scattering coefficient $\boldsymbol{\sigma}^{CG}$ of a unit cell within the CG can be obtained. It is expected that more prosperous scattering behavior can be launched in CG, because the extra coupling effect happens inside the composite lattice.

The totally transmission and reflection coefficients of *m*-th order diffraction harmonic $T_m$ and $R_m$ under Cartesian coordinate are expressed by

$$T_m = \left| \delta_{m0} + \frac{2}{L\sqrt{k_{y0}k_{ym}}} \sum_{l=-\infty}^{\infty} \left( -\frac{k_{ym}+ik_{xm}}{k_0} \right)^l \sigma_m^{SG} \right|^2 \tag{8a}$$

$$R_m = \left| \frac{2}{L\sqrt{k_{y0}k_{ym}}} \sum_{l=-\infty}^{\infty} \left( \frac{k_{ym}-ik_{xm}}{k_0} \right)^l \sigma_m^{SG} \right|^2 \tag{8b}$$

where $k_{xm} = k_0 \cos\theta + m\frac{2\pi}{L}$ and $k_{ym} = \sqrt{k_0^2 - k_{xm}^2}$. The angular radiation pattern of the scattering field is obtained by

$$\Phi(\varphi) = \left| \sum_{m=-\infty}^{\infty} \sigma_m^{SG} e^{im\left(\frac{\pi}{2}+\varphi\right)} \right|^2 \tag{9}$$

The $\sigma_m^{SG}$ in equations (8) and (9) can be replaced by $\sigma_m^{CG}$ or $\sigma_m$ to calculated the corresponding results of the CG or an isolated rod, respectively.

As mentioned above, to realize abnormal wave control with extra parameters, the dielectric grating must work in the metagrating regime $L > \lambda/(1+|\cos\theta|)$. Under this condition, the incident energy is distributed into serval discrete channels due to the diffraction effect. High-efficiency beam steering happens when the asymmetrical pattern formed by scattering of multipoles suppresses all the unselected channels. Compared with previous pioneering works[47-52] dominantly based on dipoles, we aim to improve the working performance and bring more possibilities by considering the lattice coupling and contributions of quadrupole and hexapole.

To show the flexibility of mie resonances interference, both the in-plane negative refraction and reflection induced by extraordinary optical diffraction are detailed studied.

In the metagrating regime, at least four possible diffraction channels will be available, i.e., $T_0$ with $\theta_{T0} = \theta$, $T_{-1}$ with $\theta_{T-1} = \arccos(-\lambda/L+ \cos(\theta))$, $R_0$ with $\theta_{R0} = -\theta$, and $R_{-1}$ with $\theta_{R-1} = -\arccos(-\lambda/L+ \cos(\theta))-\pi$. We firstly consider a SG with $L = 4.3r$ illuminated by a TM-polarized wave at $\lambda = \sqrt{2}L$, the parameters of which are similar to the references 50 and 52. The calculated distributions of $|E|$ are plotted in **Figure** 2a1, showing that the channels $R_0$ and $T_0$ are totally suppressed. About 20% and 79% of the input energy are distributed into the $R_{-1}$ and $T_{-1}$ channels, resulting in the in-plane negative refraction with a 90º bending angle. The multipolar coefficients of a rod inside this SG and an isolated rod (IR) are depicted in Figure 2a2 by red and blue colors, respectively. It is noticed that the lattice coupling makes the situation of a rod in SG different with the isolated case. The scattering harmonics with opposite sign no longer behave as a whole and the coupling effect among all the multipoles will change the exciting amplitude and phase.[55] Clearly, only dipoles are dominantly launched and the features of ED are observed by the near-field distribution of $|H|$ in Figure 2a3. Similar to Kerker's-type scattering, the interference between dipoloes can weak the backward propagating, which is also visualized by the far-field angular scattering patterns in Figure 2a4. The cancellation of $T_0$ is caused by the destructive interference between the incident waves and scattering waves from the cylinders. The angular scattering pattern of each rod is accompanied by two approximations, i.e., the IR and dipoles (ED+HD) approximation to see the contributions of high-order multipoles and lattice coupling. The importance of lattice coupling is obvious as the unique scattering behavior is disappearing under the IR approximation. Here, we want to point out that dipoles approximation (DA) can well reproduce the scattering pattern indicating the high-order multipoles do not participate this steering process. To enhance the working performance, one can optimize the exciting state of dipoles or take high-order multipoles into consideration, but the limited tuning freedoms of SG constrains the further improvement. Using the CG configuration can efficiently meditate this diploma due to its enriched coupling phenomenon caused by doublet. The first chosen CG (named CG1) is also taken from reference 50, and the parameters of which are $r_1 = r$, $r_2 = 0.4r$, $d = r_1+r_2+0.05r$, $L = 4.3r$ and $\phi = 90°$. The related results are illustrated in Figure 2b1-b4, indicating a working efficiency of 92%. Unlike the analysis in reference 50, it is found that this doublet dimer mainly contributes to enhancing the coefficient of MD rather than ED. Compared with the SG, $R_{-1}$ channel is further suppressed resulting from a better interference between dipoles. This improvement is induced by the interaction inside the doublet, as shown in Figure 2b3, but the high-order resonances are still lacked. Similar with above, the far-field scattering pattern of DA has tiny difference with the CG1.

To take the high-order multipoles into this process, one can increase the effective size of the composite unit, which is achieved by increasing the radius of each cylinder or enlarging the separation between them (see section 2 of the supplement). After optimizations, a new CG (named CG2) structure is chosen with parameters $r_1 = 1.06r$, $r_2 = 0.64r$, $\phi = 90°$ and $d = r_1+r_2+0.2r$. As illustrated in Figure 2c1, more than 99% of the incident energy is transferred into the $T_{-1}$ channel. It should be pointed out that high-efficiency negative refraction with a bending angle of 90° is hard to realize for common gradient metasurfaces, verifying the advantages of CG in large-angle steering. The near-field distribution in Figure 2c3 shows that strong ED and MD are excited in the bigger and smaller rods, unlike Figure 2b3 where most energy is confined in the bigger rods. Most importantly, as shown in Figure 2c2, the coupling inside the doublet will take EQ into the interference process, therefore efficiently changing the scattering angular pattern and enhancing the working efficiency. Compared with the previous two structures, the angular pattern of CG2 has a higher directionality which cannot be constructed by the approximation of dipoles. Meanwhile, all the backwards scattering channels are totally cancelled leading to an almost perfect directional scattering. The enhanced scattering phenomenon can also be explained by the general Kerker condition. Therefore, the EQ induced by lattice coupling is capable of improving the working performances. In addition, we also note that the scattering process of CG cannot be approximated by an isolated rod owing to the complex coupling process inside and among the composite lattices. To further show the promising ability of high-order multipoles, we present an extra example (CG3) with $r_1 = 1.3r$, $r_2 = 0.71r$, $\phi = 90º$ and $d = r_1+r_2+0.08r$, whose working efficiency also exceeds 99%. The corresponding results are illustrated in Figure 2d1-d4. In that case, the intriguing negative refraction is caused by the combination of first four scattering modes. In contrast to the former designs, the high-order multipolar modes play a dominant role as shown by the exciting coefficients and far-field angular scattering patterns. From the near-field map of the lattice, we also see features of high-order multipoles. This gives a clear evidence that high-order multipoles provide extra freedoms for scattering controlling, leading to the enhancement of working efficiency. Besides, utilizing the coupling inside and among lattices, the high-order multipoles which are usually hard to be controlled can be efficiently excited and applied for wavefront engineering with impressive performances.

Next, we investigate the dependence of in-plane negative refraction on wavelength and period of the grating, the results of which are plotted in **Figure** 3a and b, respectively. The theoretical calculations are accompanied by the simulations which are performed by COMSOL Multiphysics. The perfect match between the theory and simulations fully verified the validity of deduction. It shown that, near the investigated wavelengths, Fano-shaped profiles can be recognized in the spectrums of CG2 and CG3, whose peak and dip are associated with constructive and destructive interference between dipoles and high-order multipoles. This further indicates that the unity working efficiency is attributed to the contributions of high-order multipoles. SG and CG1 have smooth wavelength responses because only dipoles are excited in both structures. In addition, the calculated results in Figure 3b reveal that the high-efficiency performance of CG2 sustains in a very large period region. According to the former analysis of metagrating, each value of period is related to a bending angle. Therefore, we can carefully modify the period to change the wavefront, thus achieving a complex wave control. As a specified realization, we design a focusing lens whose working schematic is illustrated in Figure 3c. The gradient phase profile is realized by this relation of period

$$L(x) = \frac{\lambda}{|\cos(\Phi(x))| + \frac{\sqrt{2}}{2}} \tag{10}$$

where

$$\Phi(x) = \arcsin\left(\frac{f}{\sqrt{f^2 + x^2 + \sqrt{2}f|x|}} \cdot \frac{\sqrt{2}}{2}\right) \tag{11}$$

By setting the focal distance $f = 12\lambda$, a numerical simulation of a Gaussian beam impinging on the grating is simulated and presented in Figure 3d. The phenomenon of bending light and focusing into a specified position is observed, with the efficiency over 85%. The FWHM of the focusing point is about $0.98\lambda$, as shown in the inset of Figure 3d. The performance of focusing can be further optimized as the equation (10) and (11) do not consider the influence of gradient period, but the achieved result is able to prove the flexibility of periodic nanoparticles. Of course, more applications based on this wavefront-shaping method can also be expected.

The in-plane negative reflection means that the incident and reflected waves are located on the same side of the normal. According to the former analysis, the scattering waves from multipolar modes must have destructive interference with incident waves in the 0th transmission channel and have constructive interference in the -1th reflection channel. Meanwhile, the scattering behaviors in specular reflection and -1th transmission should be suppressed. It is easy to see that these requirements cannot be simultaneously met by just using the interaction between MD and ED. An optimized SG for negative reflection is set as $L = 3.1r$ and $\lambda = \sqrt{2}L$. The simulated incident, scattering, full-field waves are illustrated in **Figure** 4a1, a2 and a3, respectively. Clearly, the radiating waves have a phase difference of $\pi$ with the incident waves in the forward direction, and backward scattering waves are in phase with the incoming waves. This scattering phenomenon cancels out the incident waves and conserves the outgoing waves to -1th reflection channel, leading to the negative refraction with 89% efficiency. The scattering behavior is mainly caused by the interference between MD and EQ, as indicated by the exciting coefficients in Figure 4a4. The features of EQ are also shown in the insert of Figure 4a3 where the field enhancement induced by EQ is clearly observed. Compared with an isolated rod, the lattice coupling significantly improves the amplitude of multipoles, resulting in the unique scattering behavior. The angular scattering pattern of this CG in Figure 4a5 shows strong suppression for unwanted scattering channels which cannot be realized by the dipoles approximation or an isolated rod. Optimized from the original SG, a CG (named CG4) with $r_1 = 0.98r$, $r_2 = 0.62r$, $\phi = 0°$, $d = r_1+r_2+0.54r$ and $L = 3.1r$ is calculated and the results are illustrated in Figure 4b1-b3. The working efficiency of CG4 reaches up to 95%, because the channels of $R_0$ and $T_0$ are further constrained by the additional rod. Here, the first four multipoles are launched in the CG4 and contribute to the wave manipulation. Compared with the SG, the dipoles in CG4 play a more important role and the high-order multipoles help to suppress unwanted channels, as indicated by the far-field angular scattering patterns.

The working performance of the chosen SG for negative reflection shows great sensitivity to the wavelength, and the corresponding working bandwidth is about 5%, as plotted in **Figure** 5a. This attributes to that the properties of high-order multipoles are inherited in the negative refraction as the contribution of ED is dominant. The fragile working state also appears in the relation with period of this grating, as shown in Figure 5b. These factors may bring great challenges in practical fabrications. However, the high sensitivities for wavelength and geometry also have great potentials in many applications,[62] such as biomedical sensing, wavelength filtering and hyperspectral imaging. As a specified example for narrow-band filters or multiplexers, the field distribution of this SG is put in Figure 5c whose calculated frequency is set as $\lambda = 4r$. More than 90% of incident energy directly passes through the structure without any diffraction. The performances of CG4 are similar to these structures for negative refraction, whose dipoles are dominant and high-order multipoles help to form the scattering pattern. Therefore, the CG4 shows tolerance with the variation of wavelength or period. By carefully tailoring the geometry, the CG4 can be applied for complex wave steering. Figure 5d depicts an example by setting $L$ as $3.5r$, the reflection of which is -53° with the efficiency of 85%. Therefore, the high-order multipoles can not only improve the working performance formed by dipoles, but also dominantly achieve unique scattering patterns which are distinctive from dipoles.

In the following, we will show that the high-efficiency negative refraction and reflection realized under TM-polarized waves are also achievable for TE polarization. Same with above, we firstly calculate a CG with $L = 4.3r$ illuminated by 45° TE-polarized waves at $\lambda = \sqrt{2}L$. The field snapshot of electric field is put in **Figure** 6a1 where the efficiency of in-plane negative refraction is less than 50%. It is found that two dipoles have similar coefficients with the TM case, and the main difference is caused by the strong excitation of MQ mode, as shown in Figure 6a2 and a4. As indicated in Figure S1, the exciting wavelengths of multipoles under TE polarization are lower than that of the TM polarization. Therefore, it is hard to get an ideal interference state with dipoles only. Then, we try to optimizing the scattering pattern by adjusting MQ. Based on this, we propose and calculate an optimized CG (named CG5) with parameters $r_1 = r$, $r_2 = 0.22r$, $\phi = 90º$, $d = r_1+r_2+0.05r$ and $L = 4.3r$. The field distribution of this CG is put in Figure 6b1, indicating a working efficiency of 95%. The related multipolar exciting coefficient, near-field distribution and angular scattering patterns of a lattice are illustrated in Figure 6b2-b4. From the calculated results, we can see that the improvement of working efficiency is attributed to the strong excitation of MQ which is originated from coupling inside the doublet. The participation of MQ has great influences on the scattering pattern, leading to cancelling out two reflection channels $R_0$ and $R_{-1}$, which cannot be represented by the dipoles or an isolated rod. To the best of our knowledge, this is the first reported efficient TE-polarized negative refraction based on the dielectric metagratings because most of the previous studies only consider the contribution of dipoles.

  Inspired by this result, we further explore the possibility of higher scattering mode MH by using a SG with $r = 1.4r_1$ ($r_1$ is taken from CG5), $L = 3.06r$. The corresponding results are depicted in Figure 6c1-c4, respectively. A negative refraction with efficiency over 89% is obtained and the contribution of hexapole appears in the near-field snapshot. Because of the multiple coupling, there are also significant ED, MD, and MQ excitation, constituting the negative refraction through constructive interference. Similar to above, the negative refraction behavior is disappearing without MH, fully verifying the inevitability of high-order multipoles. These designs give a strong evidence that more prospectively controlling waves can be induced by the higher-order scattering multipoles. We also calculated working efficiency versus the working wavelength, the results of which are plotted in Figure 6d. Here, we want to emphasize that, although the quadrupole participates in the negative refraction under both TM and TE polarizations, the situations are quite different. For the results in Figure 2c, the quadrupole is applied to enhance the working performance induced by dipoles. For TE polarization, the role of quadrupole is dominant for the wave control. The similar conclusion is also valid for MH in Figure 6c. Therefore, both structures show high sensitivity to the working parameters. The relative working bandwidths are 7% and 6% for the wire gratings in Figure 6b and c.

The principle for designing negative reflection under TE polarization is similar to the previous analysis, i.e., the high-order multipoles should be taken into account. Based on this, two optimized CGs with $r_1 = 0.22r$, $r_2 = r$, $\phi = 90º$, $d = r_1+r_2+0.05r$, $L = 4.3r$ (CG6) and $r_1 = 0.42r$, $r_2 = 0.86r$, $\phi = 0º$, $d = r_1+r_2+0.44r$, $L = 4.3r$ (CG7) are chosen, and the efficiencies of them are respectively 92% and 94%. Interestingly, the parameters of CG6 are the same to that in Figure 6b when the incident angle is set as 75º. The calculated results of this CG are put in **Figure** 7a1-a4 with $\theta = 45º$ and $\lambda = \sqrt{2}L$. As indicted in Figure 7a3, the exciting coefficient of MQ is much higher than other multipoles, fully verifying its importance. Compared with the negative refraction, the phase and amplitude of the multipoles for negative reflection have been changed owing to the variation of incident angle. The different interference states contribute to changing the far-field scattering pattern, resulting in the conversion between the negative refraction and negative reflection. When it comes to CG7, the calculated results are put in Figure 7b1-b4. It is found that the features of high-order multipoles disappear as the dominant scattering modes become dipoles. Under this situation, MQ is still very important owing to its ability in suppressing $R_0$ and $T_{-1}$ channels, as indicated by the scattering patterns. The frequency responses and the dependence on period are plotted in Figure 7c and d. Consistent with the analysis, CG7 shows stability with the change of parameters, while the CG6 shows great sensibility.

**DISCUSSION AND CONCLUSION**

Note that, although the promising wave steering is realized by using the metagrating profiles, the ability of dielectric rods is obvious not limited to the serval demonstrated functions. According to the requirements, the rod grating can work in the transmission or reflection mode and the working performances can be sensitive or robust to the parameters. The multipolar interference and collective lattice coupling provide flexible and complex wavefront control for TM and TE polarizations. The working efficiency and fabrication demand of the dielectric rods can exceed the conventional gradient metasurfaces especially bending light with large angle. Note that, with the assist of asymmetrical CG, the abnormal control of waves under normal incident is also realizable. We also present the CG8 with $r_1 = 0.9r$, $r_2 = 0.58r$, $\phi = 67°$, $d = r_1+r_2+0.05r$, $L = 7r$ and CG9 with $r_1 = 0.81r$, $r_2 = 0.54r$, $\phi = 67°$, $d = r_1+r_2+r$, $L = 7r$ to realize abnormal refraction of 50° for TM and TE cases under $\theta = 90°$. The calculated wavelength is $L\sin(50°)$, leading to six diffraction channel $R_0$, $R_{-1}$, $R_1$, $T_0$, $T_1$, and $T_{-1}$. The achieved efficiencies for two structures are 92% and 90%, respectively, as shown in **Figure** 8a and b. For this design, the scattering patterns should have high directionality to suppress four unwanted diffractions and cancel out the 0th transmission, which is more complex than previous investigations. The far-field angular scattering patterns are calculated and plotted in related figures. From these patterns, we conclude all the backward scattering behaviors are well suppressed and the forward scattering is asymmetrical and highly directional, consistent with our predication. The multiple coupling inside and among the lattice helps to efficiently launch first three multipoles, then forming the desired scattering properties.

Considering the practical realization, the full-wave three-dimensional simulations are performed to study the working performance. As shown in Figure 8c, an individual lattice is chosen for calculation and periodic conditions are applied in the x direction to mimic dielectric gratings. Here, the height of rods is set as 4μm, which is realizable by conventional microfabrication process. The substrate is set as the silica with refractive index 1.45 and more details about the simulations are put in section 3 of the supplement. The simulated results of CG2, CG4, CG5 and CG7 are illustrated in Figure 8d-g, respectively corresponding to the negative refraction and reflection under both polarizations. Besides, other mentioned structures in Figure 2-8 are also calculated and put in the supplement. It is clear that, all the figures in Figure 8 are well consistent with the corresponding two-dimensional results. Therefore, although the working properties of multipoles are influenced by the height of rod[33], the manipulation of light with ultralarge angle can be well achieved under realizable situations. The parameters of gratings can be further optimized to compensate the influence of finite thickness. In addition, the influence loss in the infrared spectrum is also detailed investigated (section 4 of supplement). It is found that although the working efficiency is slightly reduced, the unique scattering patterns originated from interference among multipole are conserved when the loss is taken into account. Meanwhile, compared with dipoles, the high order multipoles show more sensitivity to the loss issues.

In conclusion, in this work we show that the high-efficiency wave steering is achievable by the interactions and coupling between multiple resonant modes of all-dielectric nanoparticles. Based on the multiple scattering theory, both the SG formed by one rod per period and the CG formed by two rods per period are explicitly analyzed under both polarizations. It is found that the strong coupling inside and among the lattice can efficiently tailor the states of multipole, leading to prosperous scattering properties. Using the simple lattice configuration at metagrating regime, the high-efficiency negative refraction and reflection with large bending angle are demonstrated under both polarizations, which is realized by controlled the diffraction channels through the interplay between multipole interferences. In addition to the fundamental dipoles, the high-order multipoles with distinctive properties are well studied and extensively applied for waves manipulation, rendering extra flexibilities for nanoparticle-based beam controls. This work can provide technique supported for on-chip optical component, including bending, sensing, filtering, focusing and so on.

## ASSOCIATED CONTENT

Supplementary material is available for some supporting information. Section 1 investigate the scattering properties of cylindrical multipoles; In Section 2, the deduction of T matrix for CG is provided and the coupling effect inside the lattice is also analyzed. Section 3 shows the configurations of three-dimensional simulations and some simulated results when the rods have finite thickness. Section 4 investigates the influence of loss on the working performance.

## AUTHOR INFORMATION

**Corresponding Author**

*Corresponding Author: pkliu@pku.edu.cn

**Author Contributions**

The manuscript was written through contributions of all authors. All authors have given approval to the final version of the manuscript. ‡These authors contributed equally. (match statement to author names with a symbol)

**Funding Sources**

National Key Research and Development Program under Grant No. 2019YFA0210203; National Natural Science Foundation of China under Grant No. 61971013.

**Notes**

The authors declare no conflict of interest.

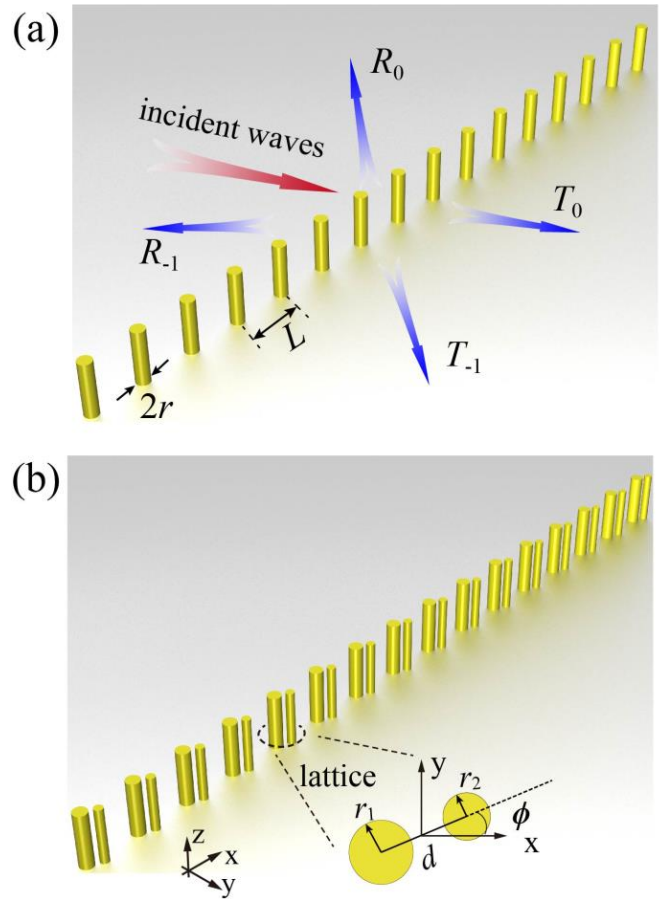

**Figure 1.** (a) shows the schematic of SG formed by a single dielectric rod in one period; (b) represents the structure of CG whose lattice consists of a doublet. All the grating is assumed to be infinite in the z direction.

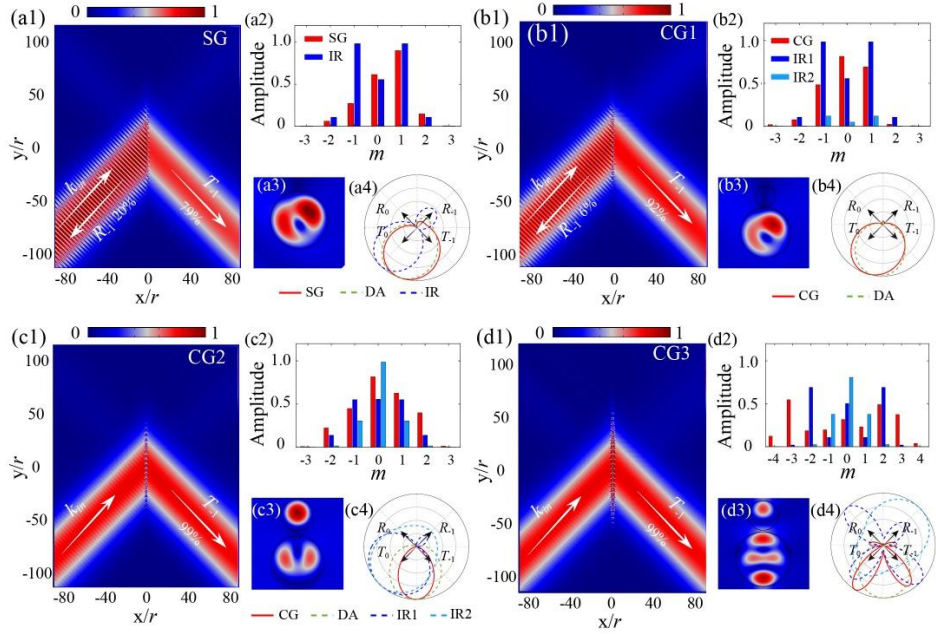

**Figure 2.** (a)-(d) illustrated the calculated results of SG, CG1, CG2 and CG3, respectively. Numbers '1', '2', '3' and '4' are respectively corresponding to the field distribution of $|E|$ when the rod array is illuminated by a Gaussian beam, the exciting coefficients of cylindrical harmonics, the near-field distribution of $|H|$ and the angular scattering patterns. The scattering pattern of grating is accompanied by the approximation of dipoles and the approximation of an isolated rod. For all the cases, the period of grating and working wavelength has the relations $\lambda=\sqrt{2}L$ and $\theta = 45°$, leading to the bending angle of $90°$.

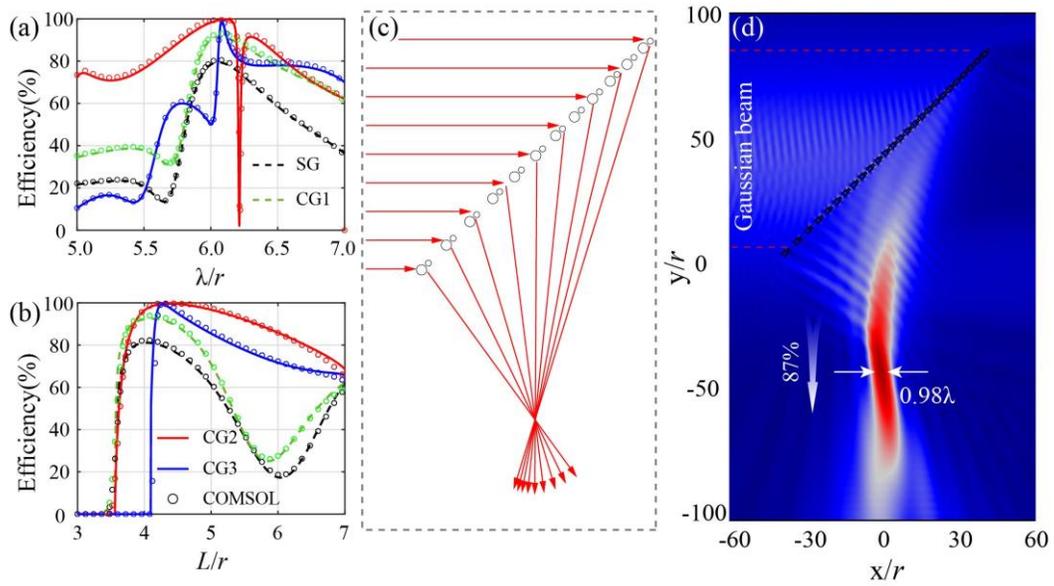

**Figure 3.** (a) and (b) show the performances of negative refraction on the dependence of working wavelength and period. The results plotted by circulars are calculated by COMSOL, which are well matched with the theory; (c) the schematic of focusing incident wave using variable rods; (d) is the field distribution of |$E$| when a Gaussian beam impinges on the deigned grating, indicating a focusing behavior with 87% efficiency.

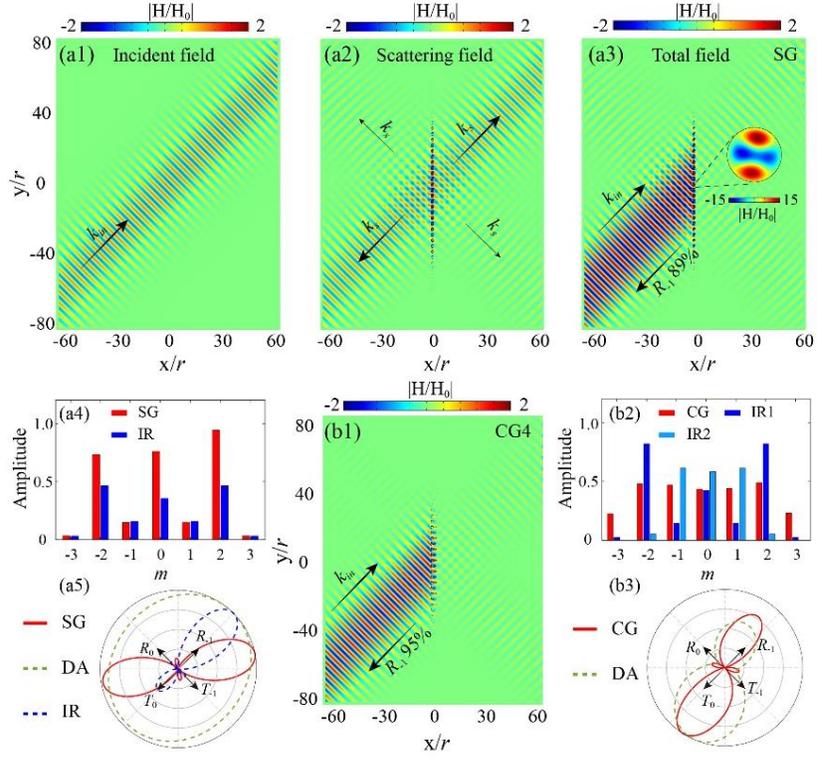

**Figure 4.** (a1)- (a3) depict the incident, scattering and total field of $H_z$, respectively. The calculated SG with $L = 3.1r$ and $\lambda = \sqrt{2}L$; (a4) and (a5) are coefficients of exciting multipoles and angular scattering patterns of this SG. (b1)- (b3) are the field map, coefficients of multipoles and angular scattering patterns of CG4.

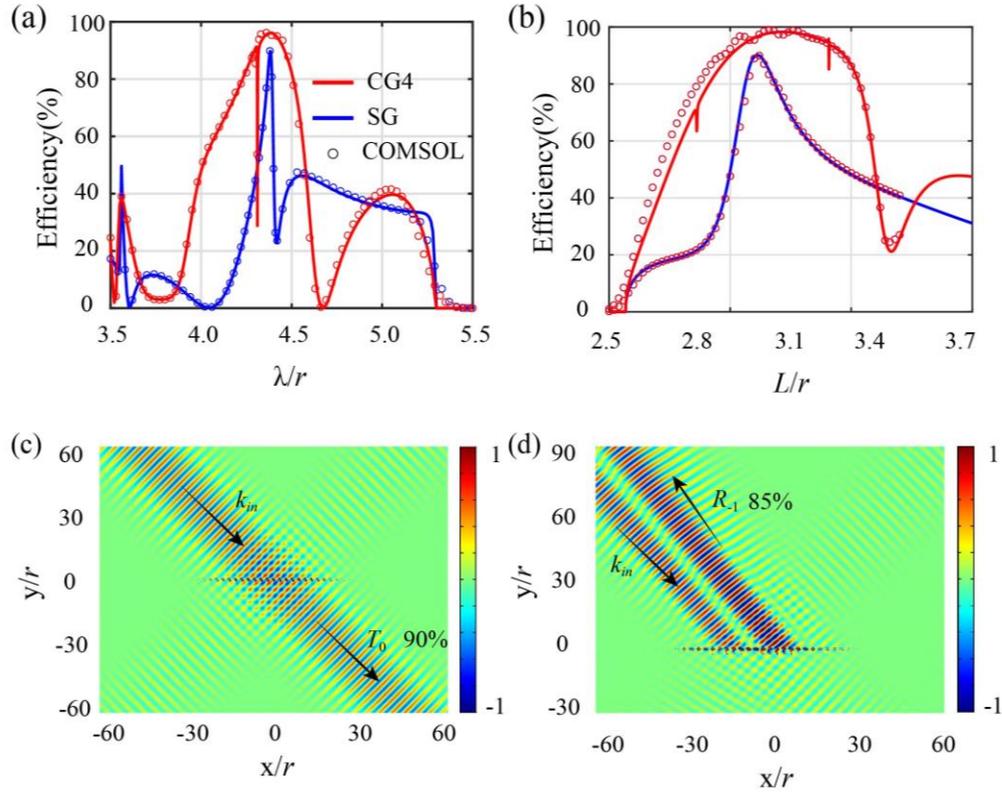

**Figure 5.** (a) and (b) are the working efficiency of negative reflection with the variation of wavelength and period of the gratings, respectively. The blue and red line are related to the theoretical results of SG and CG4, while circulars represent the simulated results by COMSOL; (c) the field snapshot of SG when the calculated wavelength is $4r$, showing that most of incident energy directly pass through the grating without any diffraction; (d) the reflection of CG4 with $L = 3.5r$.

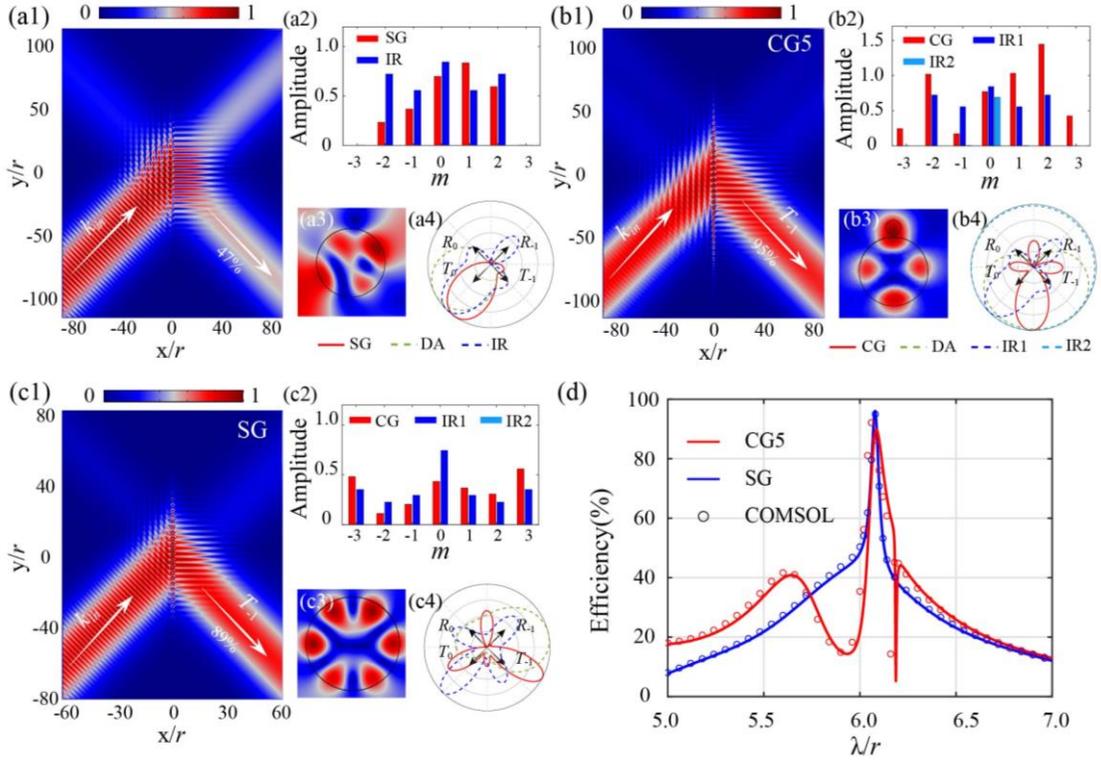

**Figure 6.** (a)-(c) show the calculated results of a reference SG with $L = 4.3r$, CG5, and an optimized SG, respectively. The numbers '1', '2', '3' and '4' are related to are respectively corresponding to the field distribution of $|E|$ when the rod array is illuminated by a Gaussian beam, the exciting coefficients of cylindrical harmonics, the near-field distribution of $|E|$ and the angular scattering patterns. The working wavelength is $\lambda=\sqrt{2}L$ and the incident angle is $\theta = 45°$, related to the negative refraction with 90° bending angle; (d) is the working efficiency of CG5 and SG versus calculated wavelength, accompanied by the simulated results by COMSOL.

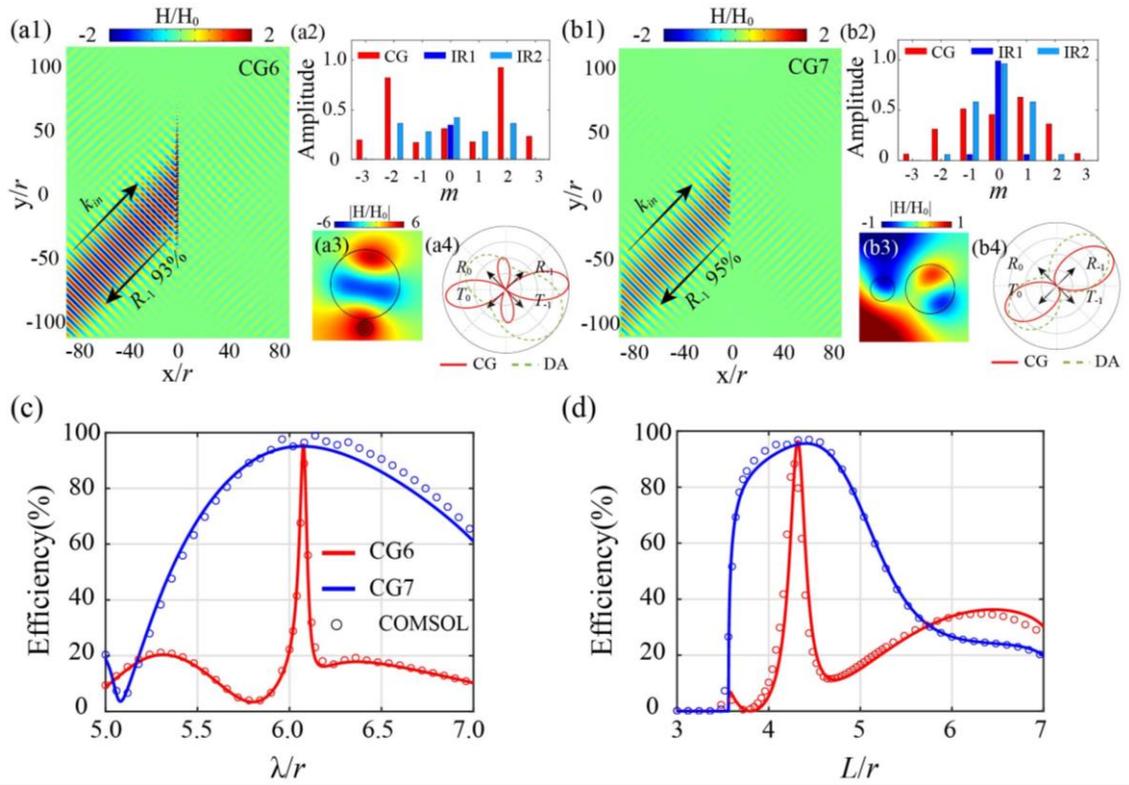

**Figure 7.** (a) and (b) show the calculated results of CG6 and CG7. The numbers '1', '2', '3' and '4' are related to are respectively corresponding to the field distribution of $E_z$ when the rod array is illuminated by a Gaussian beam, the exciting coefficients of cylindrical harmonics, the near-field distribution of $|E|$ and the angular scattering patterns. Other parameters are set as $\lambda=\sqrt{2}L$ and $\theta = 45°$; (c) and (d) are the working efficiency of CG6 and CG7 versus working wavelength and grating period, respectively.

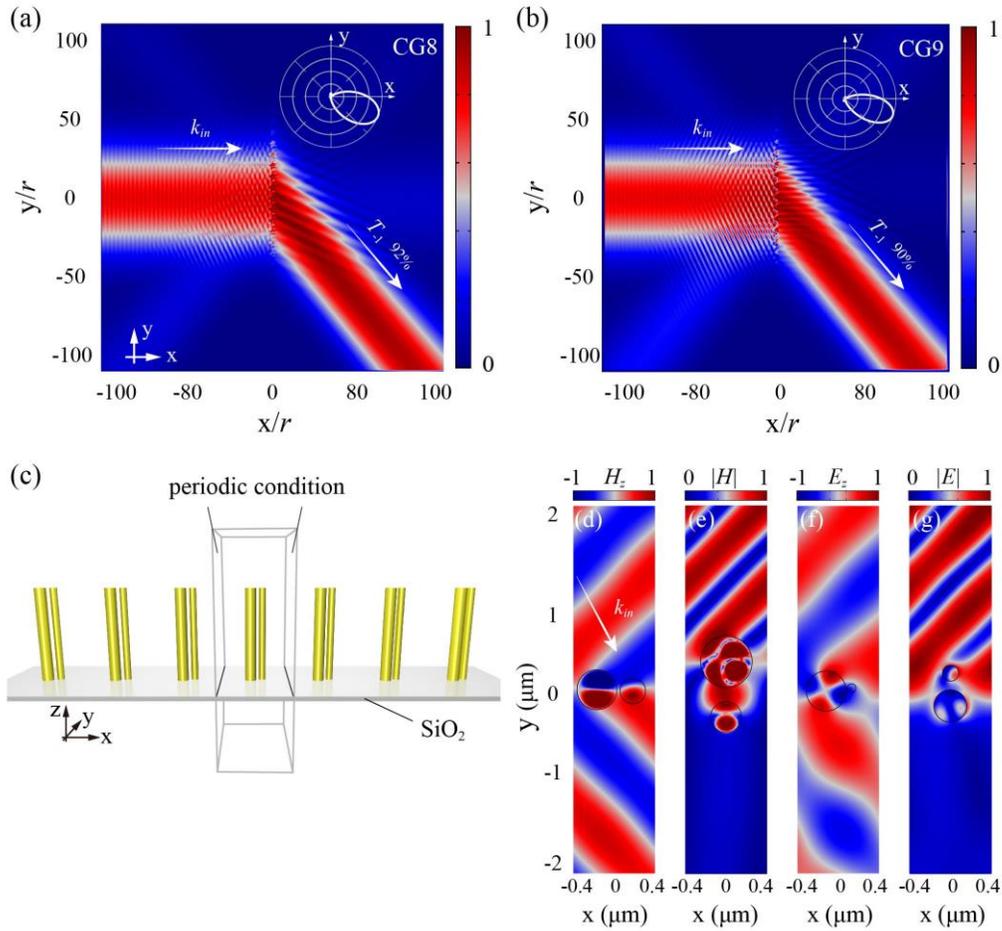

**Figure 8.** (a) and (b) are the calculated field maps of CG8 and CG9 under TM and TE polarizations, respectively. The pictures in the inset are their corresponding scattering patterns of each lattice; (c) The scheme of of simulated model when the dielectric grating has finite thickness and a silica substrate; (d)-(g) the simulated field distributions of CG2, CG4, CG5 and CG7.

36